\documentclass[journal,twoside,print]{IEEEtran}
\usepackage{tmi}
\usepackage{cite}
\usepackage{amsmath,amssymb,amsfonts}
\usepackage{algorithmic}
\usepackage{graphicx}
\usepackage{epstopdf}
\usepackage{bm}
\usepackage{siunitx}
\usepackage{textcomp}

\usepackage[acronym,toc,style=tree]{glossaries}
\newacronym{am}{AM}{Alvarez-Macovski}
\newacronym{ap}{AP}{anteroposterior}
\newacronym{ct}{CT}{computed tomography}
\newacronym{dect}{DECT}{dual-energy CT}
\newacronym[plural=dexels,firstplural=detector pixels (dexels)]{dexel}{dexel}{detector pixel}
\newacronym{fbp}{FBP}{filtered back projection}
\newacronym{fwhm}{FWHM}{full width at half maximum}
\newacronym{ha}{HA}{hydroxyapatite}
\newacronym{imbl}{IMBL}{Imaging and Medical Beamline}
\newacronym{kn}{KN}{Klein Nishina}
\newacronym[longplural={point spread functions}]{psf}{PSF}{point spread function}
\newacronym{roi}{ROI}{region of interest}
\newacronym{sirt}{SIRT}{simultaneous iterative reconstruction technique}
\newacronym{snr}{SNR}{signal-to-noise ratio}
\newacronym{spect}{SPECT}{Single-Photon Emission Computed Tomography}

\makeatletter
\def\ps@IEEEtitlepagestyle{%
	\def\@oddfoot{\mycopyrightnotice}%
	\def\@oddhead{\hbox{}\@IEEEheaderstyle\leftmark\hfil\thepage}\relax
	\def\@evenhead{\@IEEEheaderstyle\thepage\hfil\leftmark\hbox{}}\relax
	\def\@evenfoot{}%
}
\def\mycopyrightnotice{%
	\begin{minipage}{\textwidth}
		\centering \scriptsize
		Copyright~\copyright~2022 IEEE. Personal use of this material is permitted. Permission from IEEE must be obtained for all other uses, in any current or future media, including\\reprinting/republishing this material for advertising or promotional purposes, creating new collective works, for resale or redistribution to servers or lists, or reuse of any copyrighted component of this work in other works.
	\end{minipage}
}
\makeatother

\makenoidxglossaries

\newcommand{\unitspace}{\thinspace}
\DeclareSIUnit\micron{\micro\metre}

\def\BibTeX{{\rm B\kern-.05em{\sc i\kern-.025em b}\kern-.08em
    T\kern-.1667em\lower.7ex\hbox{E}\kern-.125emX}}
\begin{document}
\title{Full field X-ray Scatter Tomography}

\mycopyrightnotice

\author{Gary Ruben, Isaac Pinar, Jeremy M. C. Brown, Florian Schaff, James A. Pollock, Kelly J. Crossley, Anton Maksimenko, Chris Hall, Daniel Hausermann, Kentaro Uesugi, and Marcus J. Kitchen

\thanks{Submitted for review on XXXX-XX-XX. ``This work was supported in part by NHMRC Development Grant 1093319 and in part by a Monash University Interdisciplinary Research Grant. We acknowledge travel funding provided by the International Synchrotron Access Program (ISAP) managed by the Australian Synchrotron, part of ANSTO, and funded by the Australian Government (ISP12221). Marcus Kitchen was supported by an ARC Future Fellowship (FT160100454).''}
\thanks{Gary Ruben, James Pollock and Marcus Kitchen are with School of Physics \& Astronomy, Monash University, VIC, Australia (e-mail: marcus.kitchen@monash.edu).}
\thanks{Isaac Pinar is with Dept of Mechanical and Aerospace Engineering, Monash University, VIC, Australia.}
\thanks{Jeremy M. C. Brown was with School of Physics \& Astronomy, Monash University. He is now with Centre for Medical Radiation Physics, University of Wollongong, Wollongong, Australia.}
\thanks{Florian Schaff was with School of Physics \& Astronomy, Monash University. He is now with Dept of Physics, Technical University Munich, Germany.}
\thanks{Kelly Crossley is with The Ritchie Centre, Hudson Institute for Medical Research, Clayton, VIC, Australia.}
\thanks{Anton Maksimenko, Chris Hall and Daniel Hausermann are with The Australian Synchrotron – ANSTO, Clayton, VIC, Australia.}
\thanks{Kentaro Uesugi is with Japan Synchrotron Radiation Research Institute (JASRI)/SPring-8, Sayo, Hyogo, 679-5198, Japan.}}

\maketitle

\begin{abstract}
In X-ray imaging, photons are transmitted through and absorbed by the target object, but are also scattered in significant quantities. Previous attempts to use scattered X-ray photons for imaging applications used pencil or fan beam illumination. Here we present 3D X-ray Scatter Tomography using full-field illumination for small-animal imaging. Synchrotron imaging experiments were performed on a phantom and the chest of a juvenile rat. Transmitted and scattered photons were simultaneously imaged with separate cameras; a scientific camera directly downstream of the sample stage, and a pixelated detector with a pinhole imaging system placed at 45${}^\circ$ to the beam axis. We obtained scatter tomogram feature fidelity sufficient for segmentation of the lungs and major airways in the rat. The image contrast in the scatter tomogram slices approached that of transmission imaging, indicating robustness to the amount of multiple scattering present in our case.
This opens the possibility of augmenting full-field 2D imaging systems with additional scatter detectors to obtain complementary modes or to improve the fidelity of existing images without additional dose, potentially leading to single-shot or reduced-angle tomography or overall dose reduction for live animal studies.

\end{abstract}

\begin{IEEEkeywords}
Compton, scatter, tomography, X-ray imaging, preclinical imaging
\end{IEEEkeywords}

\section{Introduction}
\label{sec:introduction}
\IEEEPARstart{A}{ttenuation-based} X-ray imaging employs only a fraction of the photons incident on an object to produce image contrast. The remainder are absorbed or scattered. For clinical X-ray imaging, the presence of scattered photons at the detector plane is typically considered a hindrance because they form a diffuse background, reducing the image contrast-to-noise ratio \cite{Men18, AlsMc11}. Anti-scatter grids are therefore often used to selectively prevent them from reaching the detector. Because photons scatter to all angles, they can be captured with additional detectors located around the object. By doing so in a way that constrains the photon path information, tomography can be performed. Our aim is to employ the scatter signal produced during X-ray imaging to obtain volumetric information. In contrast with previous studies, which took non-full-field approaches, we augment an existing X-ray imaging system used for preclinical studies of lung function in small animals, by adding a detector to collect these unused photons. Combining additional views from multiple detectors placed around the object would then open the possibility of doing single-shot, time-resolved, or reduced-rotation-angle tomography. The purpose of this study is to establish the feasibility of scatter tomography for full-field imaging of small animals.

The use of scattered photons has a long history in X- and $\gamma$-ray imaging. Lale \cite{Lal59} is considered the earliest to propose Compton scattering for medical imaging using a pencil beam X-ray source. 
Following Lale's work, Compton imaging was pursued as a medical imaging approach for application to radiotherapy treatment planning \cite{FarCo71, FarCo74}, bone densitometry \cite{HudBh79a}, lung function measurement \cite{KauGa76, GarWe77, DelBe82}, and heart motion tracking \cite{HerMc94}.
Notably, Farmer and Collins identified the problem of photon self-absorption \cite{FarCo74}, which causes artifacts and departure from quantitative correctness when applying conventional tomographic reconstruction approaches. For a review of the theory and history of the medical application of scatter imaging, covering the period starting from Lale's work through to the late 1990s, see Harding \cite{Har97}.

Scatter imaging also allows for imaging geometries where the source and detector are both on the same side of an object. This has been useful for imaging large objects, such as in airport baggage scanning, integrity testing of assembled aircraft, and cultural heritage applications like scanning ground or walls \cite{HarHa10}. In such scenarios, scatter imaging provides information where transmission imaging cannot.

Earlier medical scatter imaging research compared the dose delivered to a human patient solely from scatter imaging to that from transmission imaging \cite{BatSa77, BatBr81}. Combining the two imaging methods could instead improve existing images while maintaining dose, or maintain image fidelity while lowering overall dose.
In cases where a radiotherapeutic dose is already being delivered to a patient, the scattered photons produced as a byproduct could be used for imaging, allowing, for example, guided therapy by using the scattered photons to perform organ imaging \cite{BroKi16, YanTi16}.

Full-field X-ray imaging employs an extended beam to simultaneously illuminate a wide region of a subject. This allows a camera to image the region of interest with a single exposure, and supports time-resolved imaging or \gls{ct} without the need to translate the subject or detector.
Scatter-based \gls{ct} can produce a map of electron density rather than the map of linear attenuation coefficient produced by conventional transmission \gls{ct}. In medical applications such as proton and heavy ion radiotherapy planning, where the electron density map is needed for simulation of the interaction of those sources with the patient, the scatter-based \gls{ct} approach provides a more direct measurement of the quantity of interest \cite{TorTs03}.

A disadvantage of full-field illumination over other geometries such as sheet or pencil-beam illumination is that additional paths exist for multiply-scattered photons to reach the detector. This is also true for the case of transmission imaging, even with anti-scatter grids. In both cases, conventional \gls{ct} reconstruction approaches cannot resolve these paths and they thus contribute to a background signal. At higher energies than those studied here, which are required for human patient imaging, the ratio of multiply-scattered to singly-scattered photons exiting the object increases. This makes full-field illumination problematic for tomography due to the increased difficulty of isolating the full photon-path history.
In small animal imaging, the object size and elemental composition involved allow lower energy X-ray illumination, below $\sim$40\unitspace\si{keV}. At these energies it becomes viable to consider scatter tomography with full-field illumination because multiply-scattered photons are reduced in energy with each Compton scattering event, typically associated with an increase in attenuation. At the same time, the additional scattering paths increase the mean total path length within the object. These effects combine to reduce the unwanted background and, as we show, make full-field scatter tomography feasible.

\section{Tomography based on scattered photons}
\label{sec:scatter_tomography}

\subsection{Physical processes}
\label{sec:physical_processes}
We now describe the processes that photons undergo from source to detector and how tomography can be used to map the electron density in an object. This description applies to illumination with a monochromatic parallel beam, typically associated with synchrotron imaging, but we foresee little impediment to its extension to lab-based X-ray sources.

Photon scattering from free and bound electrons is described, respectively, by the Compton (inelastic) and Rayleigh (elastic) scattering processes. Rayleigh scattering is mostly forward-directed and decreases with increasing scattering angle, transitioning to where Compton scattering dominates. All mentions of `scattered photons' in this manuscript refer to the combination of these processes.

In parallel-beam transmission \gls{ct}, attenuation of X-rays by an object in the direction of propagation is responsible for the contrast in images acquired by a transmission detector. The inverse Radon transform is applied to perform tomography on a rotation sequence of such images \cite{KakSl88}. Single-scattering scatter \gls{ct} differs primarily in that the photon path is composed of two subpaths, with a single scattering site at the nexus, and only the first part of the path lies along the primary beam axis. X-ray photons enter, interact with, then exit the object before finally reaching the detector. The X-rays may undergo photoelectric absorption while traveling along the entry and exit paths within and outside the object. The inverse Radon transform assumes attenuation occurs through the full extent of the object and hence does not apply directly to scatter tomography. If the exit subpath can be determined, such as by collimation, the knowledge of the photon directions allows the inverse Radon transform to be applied, albeit with incorrect signal values. Unless the attenuation is numerically corrected, this results in a reconstruction that exhibits self-absorption artifacts.

\subsection{Interaction Model}
\label{sec:interaction_model}

\begin{figure}
	\centering
	\includegraphics[width=0.6\columnwidth]{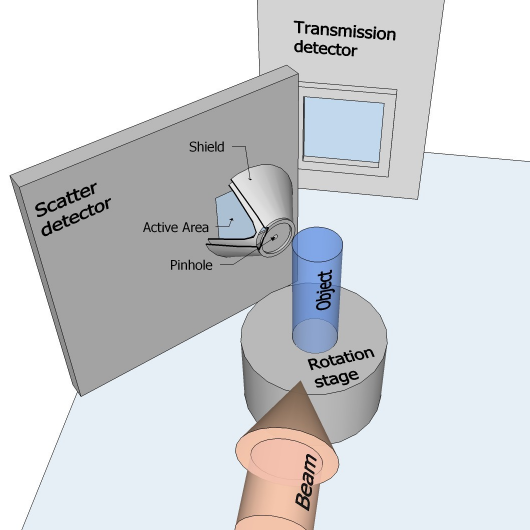}
	\caption{\label{fig:experimental_geometry} Experimental arrangement. A full-field monochromatic beam illuminates an object on a rotation stage. Transmission and scatter images are acquired simultaneously with pixel-based detectors placed on-axis (Transmission detector) and at 45${}^\circ$ to the beam axis (Scatter detector), with the latter located so as not to impede the direct beam path. Scattered photons are imaged using pinhole collimation.}
\end{figure}

An illustration of our experimental setup is shown in Fig.~\ref{fig:experimental_geometry}.
Starting with a conventional parallel-beam geometry, a scatter detector integrated with a pinhole collimator was added. Its placement was both close to the object and close in angle to the forward beam axis, while avoiding impeding the beam path to the transmission detector.

\begin{figure}
	\centering
	\includegraphics{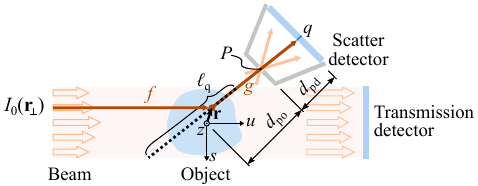}
	\caption{\label{fig:beamlet_model} Interaction model. An isolated beamlet with incident intensity $I_0(\mathbf{r}_\perp)$ illuminates a row of the object. It is attenuated according to $f$ before scattering from a site with coordinate vector $\mathbf{r}$. The resulting scatter is further attenuated according to $g$, along the path passing through the pinhole with fixed coordinate $P$ before being detected by dexel $q$. As different dexels $q$ are considered, the outgoing ray pivots around $P$, and line segment $\overline{Pq}$ constrains the outgoing path and all points and line segments that lie along it. The total intensity measured at $q$ considers all contributions from scattering along the dotted line segment $\ell_\mathrm{q}$ whose length is limited by the width of the beam. The pinhole-image backprojection algorithm makes use of the distances from the pinhole to the detector $d_\mathrm{pd}$ and to the object $d_\mathrm{po}$.}
\end{figure}
We describe the interaction using a single-scattering model, referring to the schematic in Fig.~\ref{fig:beamlet_model}.
When considering full-field illumination of the object, the incident X-ray beam can be treated as a collection of independent beamlets, each illuminating a row of scattering sites parallel to the $u$ axis.

We start by considering a single scattering site $\mathbf{r}$ which lies on a straight line constrained by the pinhole center and a \gls{dexel} identified by integer index $q$. An incident beamlet with intensity $I_0(\mathbf{r}_\perp)$ intersects the object at transverse position $\mathbf{r}_\perp \equiv (r_\mathrm{s}, r_\mathrm{z})$, where $\mathbf{r} \equiv (r_\mathrm{s}, r_\mathrm{u}, r_\mathrm{z})$ is the scattering-site position-vector, in terms of its Cartesian components.
The incident X-rays are attenuated according to a function $f$ before scattering with probability $p$ from the site $\mathbf{r}$ towards the pinhole. If absorption by air is significant, the segments of both the incoming and outgoing paths outside the object should be included. The scattering occurs via a combination of the Compton and Rayleigh processes. For the Compton scattering component, there is an associated energy reduction. After scattering at $\mathbf{r}$, the X-rays are further attenuated, according to a function $g$, before being detected at $q$. The beamlet's contribution to the measured intensity at $q$ is
\begin{equation}
	\mathrm{d}I(\mathbf{r}_\perp, q) = I_0(\mathbf{r}_\perp) f(\mathbf{r}) p(\mathbf{r}) g(\mathbf{r}) \,\mathrm{d}\ell
	\label{eq:dexel_intensity_from_Isz},
\end{equation}
where $\mathrm{d}\ell$ is an infinitesimal element of the outgoing path.
The intensity also depends on the object and \gls{ct}-stage orientation $\Theta$. For a single image, $\Theta$ is constant and we omit it for clarity.

The incident beamlet is attenuated according to the well-known Beer-Lambert law
\begin{equation}
f(\mathbf{r}) = \exp\left( -\int_{u_\mathrm{src}}^\mathbf{r} \mu_\mathrm{B}(\mathbf{r}_\perp, u) \,\mathrm{d}u \right).
\label{eq:inpath}
\end{equation}
The linear attenuation coefficient $\mu_\mathrm{B}$ is that of the particular material in the differential element $\mathrm{d}u$, which is determined at the incident monoenergetic beam energy $E_0$ for our case of synchrotron radiation. The proportion of photons at site $\mathbf{r}$ scattered into the outgoing path that will reach the \gls{dexel} $q$ is
\begin{equation}
p(\mathbf{r}) = \rho_\mathrm{e}(\mathbf{r}) \cdot \, \dfrac{1}{4\pi} \int_{\Omega_\mathrm{q}} \!\! \dfrac{\mathrm{d}\sigma}{\mathrm{d}\Omega}\Big(C_m, E_0, \theta, \phi \Big) \, \mathrm{d}\Omega,
\label{eq:scattering_probability}
\end{equation}
where $\rho_\mathrm{e}$ is the electron density, $C_m$ is the material compound, and $\theta$ and $\phi$ are the polar-coordinate scattering angles toward $q$, all specific to the scattering site. The Compton and Rayleigh scattering patterns depend on $C_m$, the incident energy $E_0$, and scattering direction $(\theta, \phi)$. The differential cross sections for these scattering types are incoherently summed to form the total differential cross section $\frac{\mathrm{d}\sigma}{\mathrm{d}\Omega} \equiv \frac{\mathrm{d}\sigma_\mathrm{C}}{\mathrm{d}\Omega} + \frac{\mathrm{d}\sigma_\mathrm{R}}{\mathrm{d}\Omega}$, with the subscripts indicating Compton and Rayleigh scattering, respectively. These cross sections depend on the material's electron configuration which is incorporated via incoherent scattering functions and atomic form factors \cite{SchBr11}. The differential cross section is integrated over a solid angle $\Omega_\mathrm{q}$ which defines the combined extents of the pinhole and dexel $q$. $\Omega_\mathrm{q}$ can also incorporate the \gls{psf} of the detector scintillator layer, provided that is also of a similar extent to the effective aperture. In this way, $\Omega_\mathrm{q}$ and the total \gls{psf} both give corresponding measures of the image feature resolution. 
Furthermore, $\Omega_\mathrm{q}$ can be treated as constant, provided that the distance from the pinhole to the closest point of the object is always large relative to the pinhole aperture.

The electron density relates to mass density $\rho$ according to $\rho_\mathrm{e}=N_a\rho Z/M$, where $N_a$ is Avogadro's number, $Z$ is the atomic number, and $M$ is the molar mass \cite{AlsMc11}. For a compound, $Z/M$ is replaced by the weighted sum over elements $\sum_i w_i(Z/M)_i$.

The photons scattered from $\mathbf{r}$ into the outgoing path are again attenuated according to the Beer-Lambert law
\begin{equation}
g(\mathbf{r}) = \exp \left( - \int_{\mathbf{r}}^{\mathrm{q}} \!\!\!\!  \mu_\mathrm{S}(l) \,\mathrm{d}l \right).
\label{eq:outpath}
\end{equation}
before reaching the \gls{dexel} $q$. The path-dependent attenuation coefficient $\mu_\mathrm{S}$ depends on the energy of the scattered photons. Rayleigh scattered photons maintain the incident energy $E_0$, whereas Compton scattered photons' energy $E'$ is reduced according to Compton's energy-angle relationship
\begin{equation}
E' = \dfrac{E_0}{1+\frac{E_0}{m_e c^2}(1-\cos \theta)},
\end{equation}
for a photon scattered from a free electron at rest through an angle $\theta$, where $m_e$ and $c$ are the electron rest mass and speed of light, respectively. The attenuation coefficient is a combination of the coefficients evaluated at each of the two energies $E'$ and $E_0$, weighted in proportion to the associated scattering type. An alternative approach can be taken where the outgoing path is treated as two coincident path branches, with each assigned one of the two energies. This would also allow accounting for the energy response of the scintillator material. A third practical approach is to assign the Compton energy $E'$ to all scattered photons, since most Rayleigh scattering is through small angles, where $E'$ is close to $E_0$.

Unlike pencil beam illumination, for full-field illumination the intensity at $q$ includes contributions from all scattering sites along a line segment $\ell_\mathrm{q}$, which is collinear with the line segment $\overline{Pq}$ passing through the pinhole center $P(s, u, z)$ and \gls{dexel} $q(s, u, z)$.
The detected intensity $I(q)$ is the sum of contributions $\mathrm{d}I(\mathbf{r}_\perp, q)$ [see (\ref{eq:dexel_intensity_from_Isz})] from all scattering sites with position $\mathbf{r}(s, u, z)$, which now varies along $\ell_\mathrm{q}$
\begin{equation}
	I(q) = \int_{\ell_\mathrm{q}} \mathrm{d}I(\mathbf{r}_\perp, q) = \int_{\ell_\mathrm{q}} I_0(\mathbf{r}_\perp) f(\mathbf{r}) p(\mathbf{r}) g(\mathbf{r}) \,\mathrm{d}\ell.
	\label{eq:dexel_intensity}
\end{equation}

The above description relates the source intensity distribution and measured intensities for every \gls{dexel} $q$ to form a complete image.
In their paper, Brunetti \textit{et al.} \cite{BruCe02} present a description of Hogan's attenuation correction approach \cite{HogGo91}, which derives an expression $\bar{K}$ for the azimuthally-averaged combination of the attenuation functions and differential cross sections, and transforms it to a form that is identifiable as the Radon transform of the product of $\bar{K}$ and the scattering probability $p(s, u, z)$. This relation is then inverted using the inverse Radon transform, numerically realized with \gls{fbp} or iterative back projection, resulting in the map of electron density $\rho_\mathrm{e}(s, u, z)$. Brunetti \textit{et al.} used illumination by a pencil beam, and a bucket detector. Due to our  significantly different forward model description, we have not attempted to pursue an analytical attenuation correction. Nor have we attempted numerical attenuation correction. Approximations employed by attenuation correction methods typically limit their application to certain types of objects and materials. Because we have a map of linear attenuation coefficient at the incident energy available, it may be possible to approximate the attenuation coefficient at the Compton energy for the materials in our samples and to iteratively compute the attenuation-corrected electron density map by following an approach similar to Golosio \textit{et al.} \cite{GolSi03}.

In Brunetti \textit{et al.} \cite{BruCe02} the detector was placed at right angles to an incident X-ray pencil beam. In that geometry, Compton scattering predominates over Rayleigh scattering, so the scattering probability (\ref{eq:scattering_probability}) is well characterized purely by a \gls{kn} cross section $\sigma_\mathrm{KN}$. In our description we allow for arbitrary detector placement and, in the experiment described here, endeavor to place the detector closer to the forward beam axis direction for two reasons. First, we use highly polarized synchrotron radiation, which results in a scattering distribution minimum on the horizontal axis at 90${}^\circ$, which would severely limit the photon flux reaching a detector at that location. We note that this Compton scatter minimum would also contribute to a relative enhancement there of any residual fluorescence photons generated by inner shell ionization that might escape the object. However, for the size and composition of our specimens, any fluorescence will be insignificant due to the low fluorescence photon energies and consequent strong self-absorption in the specimen.
Second, both Rayleigh and Compton scattering are enhanced forward of the scatter site, so placing a detector closer to the beam axis should capture more photons.
Physical constraints make it challenging to place the detector to capture Rayleigh scattered photons, so their contribution may be ignored for detectors placed far from the forward beam axis.

Because we use a pinhole to collimate or define the photon paths between the object and detector, the rays reaching the detector are those that converge at the pinhole, forming a cone. Tomographic reconstruction is therefore performed using cone beam back projection. However, compared with the conventional cone beam case, the object is on the opposite side of the apex and its images are inverted, so the distances usually used to define the geometry are modified accordingly.

\section{Materials and Methods}
\subsection{Materials}
\begin{figure}
	\includegraphics[width=\columnwidth]{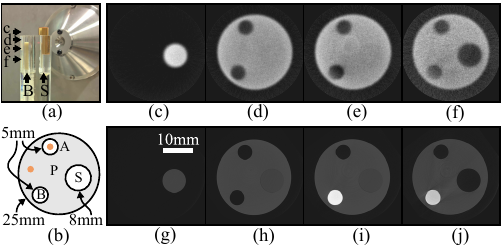}
	\caption{\label{fig:phantom} Scatter and transmission tomographic imaging of a cylindrical PMMA phantom with cavities containing air, bone-equivalent plastic [substrate with hydroxyapatite (HA)], and the same substrate without HA. (a)~View of phantom towards the scatter detector with conical pinhole holder and detector shielding visible. The phantom was illuminated by X-rays from the left and rotated about its vertical axis. The inserts were vertically offset to create different composition combinations in the slices. (b)~Phantom composition reference: bone-equivalent plastic B and substrate S, both indicated in (a), air A, and PMMA P. Two regions of interest shown in orange were used to calculate image contrast. (c--f)~Scatter tomogram slices, each averaged from 61 slices at locations indicated in (a). The electron density $\rho_\mathrm{e}$ values are shown in arbitrary units, with black representing 0. (g--j)~Corresponding transmission tomogram slices showing linear attenuation coefficient $\mu$ on a grayscale map range from 0 (black) to 1.7\unitspace\si{\per\cm\tothe{1}} (white).}
\end{figure}

Our imaging subjects were a plastic (PMMA) cylindrical phantom [Fig. \ref{fig:phantom}(a--b)] and a deceased juvenile rat scavenged from a previous terminal experiment with approval from the Australian Synchrotron and Monash University animal ethics committees.

The PMMA phantom has three cavities. One was left empty and open to air, one contained a bone-equivalent insert composed of a plastic substrate with 750\unitspace\si{mg\per\cm\tothe{3}} \gls{ha}, and one contained an insert of the same substrate, without \gls{ha}.
The \gls{ct} arrangement was quite conventional, with each specimen centered on a rotation stage, and fully illuminated by a monoenergetic beam (Fig.~\ref{fig:beamlet_model}). A transmission detector [pco.edge 55 sCMOS camera with 25\unitspace\si{\micron} thick Hamamatsu P43 (Gd\textsubscript{2}O\textsubscript{2}S:Tb) scintillator and tandem lens system] was placed 58\unitspace\si{cm} downstream of the specimen. The scatter detector (Hamamatsu C9728DK-10) was a CMOS detector with 50\unitspace\si{\micron} pixels and an integrated CsI scintillator directly coupled to the sensor. This was placed at 45${}^\circ$ to the beam axis and 51\unitspace\si{mm} from the rotation axis, to be as close as practical to the specimen and to the forward beam direction whilst avoiding impeding the beam path to the transmission detector. A circular pinhole with diameter $0.56 \pm 0.01$\unitspace\si{mm} in a 0.6\unitspace\si{mm} thick lead sheet was held near the apex of a conical 5\unitspace\si{mm} thick stainless steel X-ray shield, and centered over the scatter detector 42\unitspace\si{mm} from the scintillator [Fig. \ref{fig:phantom}(a)].

\subsection{Method}
Experiments were performed at the Australian Synchrotron \gls{imbl} in hutch 3B \cite{SteCr17}.
The beamline wiggler field was set to 2T to minimize higher-energy harmonics that might produce photons at undesirable energies via single or multiple Compton scattering.
The choice of energy in any X-ray imaging system is governed by the subject's material composition and length scales. For our small animal and phantom, good image contrast was achieved with an incident beam energy of 30\unitspace\si{keV}. At this energy, photons scatter via the Compton and Rayleigh processes in similar proportions, although their angular distributions differ.

A summary of the experimental geometry parameters is provided in Table~\ref{tab:geometric_parameters}.
\begin{table}
	\caption{Experimental geometry parameters}
	\centering
	\begin{tabular}{crl}
		& Parameter & Value\\
		\hline
		Global & Source-object distance & $\sim$140\unitspace\si{\m} \cite{SteCr17} \\
		\\
		& Propagation distance & 580\unitspace\si{\mm} \\
		Transmission & Square dexel side & 12.5\unitspace\si{\micron} \\
		& Detector active area & 2160 rows $\times$ 2560 cols \\
		\\
		& Pinhole diameter & $0.56 \pm 0.01$\unitspace\si{mm} \\
		& $d_\mathrm{po}$ (Fig.~\ref{fig:beamlet_model}) & 53.6\unitspace\si{mm} \\
		& $d_\mathrm{pd}$ (Fig.~\ref{fig:beamlet_model}) & 42\unitspace\si{mm} \\
		Scatter & Beam-axis to detector angle & 45\unitspace\si{\degree} \\
		& Square dexel side $\delta_x$ & 50\unitspace\si{\micron} \\
		& Detector active area & 1032 rows $\times$ 1032 cols \\
	\end{tabular}
	\label{tab:geometric_parameters}
\end{table}
The rat was held using cling film in a PMMA semi-annular holder, with its anterior-posterior axis held vertical, close to the rotation axis. The lungs were aerated with a custom-designed ventilator \cite{KitHa10} via an endotracheal tube that was sealed off at a fixed pressure of 20\unitspace\si{cm}H\textsubscript{2}O.
For each specimen, 1,600 scatter-image exposures of 1/3\unitspace\si{s} were acquired at equispaced angles over 360${}^\circ$. Scatter images were acquired with the software \si{\micro}Manager \cite{EdeTs14}. Because of limitations in our timing source, three transmission images were acquired for each scatter image, i.e. 4,800 40\unitspace\si{ms} exposures. The same exposure timings were used to acquire 50 dark-field and 50 flat-field images for correcting the transmission images and performing \gls{ct} reconstruction.

To achieve quantitative results in synchrotron transmission imaging, the source beam's nonuniform intensity profile must be accounted for. A flat-field image of the beam captures the spatial intensity distribution, which characterizes the source for all subsequently measured images. Conventional \gls{ct} reconstructs the attenuation coefficient from the ratio of measured intensities with and without the target object in place, with the flat-field image providing the latter measurement.

\begin{figure}
	\centering
	\includegraphics[width=\columnwidth]{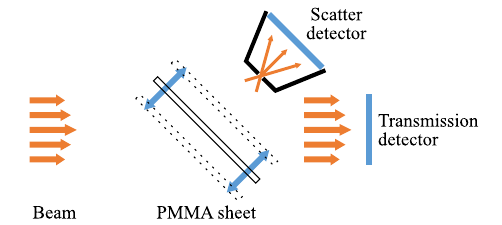}
	\caption{Flat-field correction was performed by replacing the specimen with a PMMA sheet. With the tomography stage static, the PMMA sheet was translated perpendicular to the scatter detector face and imaged at different locations throughout the volume swept out by the specimen. Averaging these scatter images produced an intensity map akin to a flat-field image, suitable for correction of the specimen scatter images.}
	\label{fig:flat_fielding}
\end{figure}

The scatter images are similarly affected by the source nonuniformity, but for these a candidate for the equivalent beam intensity profile is not directly available. To create a relevant profile, we used a uniform scattering object –- a PMMA sheet acting as a probe –- as a proxy for the source, generating a secondary source of photons that emits toward the scatter detector in proportion to the primary beam intensity as it intersects and scatters from each location in the probe (Fig.~\ref{fig:flat_fielding}).
A 6\unitspace\si{mm} thick PMMA sheet was used, as it is thick enough to generate a sufficient scattered photon signal, yet thin enough to minimize self-absorption and multiple-scattering effects.
The sheet was shifted perpendicular to the scatter detector face and imaged at different positions that cover the volume defined by the target object during its imaging sequence. Shifting the probe accounts for two changes that occur with distance from the pinhole; magnification, and a horizontal shift of the intensity peak. The profiles, each an average of 50 1/3\unitspace\si{s} exposures, were further averaged to produce a single flat field image $I_\mathrm{F}$.
This method is an approximation, because any given orientation of the object only scatters from a subset of the volume probed by the sheet.

A dark field image $I_\mathrm{D}$ was obtained by averaging 50 1/3\unitspace\si{s} exposures taken with the beam off.
Each object scatter image $I_\mathrm{S}(\Theta)$ was divided by $I_\mathrm{F}$ in a conventional flat-fielding scheme to produce scatter images $I(\Theta) = [I_\mathrm{S}(\Theta)-I_\mathrm{D}]/[I_\mathrm{F}-I_\mathrm{D}]$ ready for backprojection.
An alternative, equivalent approach is to generate a rotationally symmetric 3D correction volume from the $I_\mathrm{F}$ images and apply this to the uncorrected tomogram.

\subsection{Transmission image data processing}
The transmission images were reconstructed using parallel-beam \gls{fbp} with the X-TRACT \cite{GurNe11} software package, which performed the flat and dark field correction and TIE-Hom single-image phase retrieval to remove propagation-based phase contrast fringes \cite{PagMa02}. We used every second image (1,200 images) over 180${}^\circ$. The phase retrieval parameters used for the rat were $R=0.50$\unitspace\si{m} and $\delta/\beta=1,980$, where $R$ is the source-detector distance and $\delta/\beta$ is the ratio of parts of the complex refractive index $n=1-\delta+i\beta$. The parameters for the PMMA phantom were $R=0.20$\unitspace\si{m} and $\delta/\beta=2,495$.

\subsection{Scatter image data processing}
We operated the detector above its upper rated energy limit of 18\unitspace\si{keV}, resulting in many direct photon detections, which appear as bright outliers (salt) in the image. If ignored, these bright pixels lead to streaks in the iteratively-reconstructed tomogram and can affect convergence of iterative reconstruction schemes. Thus, they were removed from the dark-field corrected images by outlier identification using a rank filter and subsequent replacement by the local-spatial-mean value.

The processed images were used to reconstruct the 3D tomogram using 100 iterations of Astra toolbox's 3D \gls{sirt} cone beam reconstruction algorithm \cite{vanPa15}. Other total iteration counts straddling the chosen value 100 were also performed, and judged to produce overly blurry or noisy reconstructions.
Conventional cone beam reconstruction depends on the source-object distance $d_\mathrm{so}$, object-detector distance $d_\mathrm{od}$, and \gls{dexel} size $\delta_x$. The pinhole imaging geometry is instead characterized by pinhole-to-detector distance $d_\mathrm{pd}=42$\unitspace\si{mm}, pinhole-to-rotation-center distance $d_\mathrm{po}=53.6$\unitspace\si{mm}, and $\delta_x$. The two descriptions are related by observing that $d_\mathrm{so}=d_\mathrm{pd}/2$ and $d_\mathrm{od} + d_\mathrm{so} = d_\mathrm{pd}$. To reconstruct the scatter image data, the distances $d_\mathrm{so}=d_\mathrm{od}=d_\mathrm{pd}/2$ were used as inputs to Astra's \gls{sirt} algorithm, resulting in a tomogram with voxel side length $\delta_x d_\mathrm{po}/d_\mathrm{pd}$.

Despite the flat field correction, a residual vertical intensity gradient remained uncorrected in the reconstructed volume. This may be because we sampled the scatter profile with the PMMA sheet at only two positions, which may have been inadequate. The scatter profile correction should work best if the sheet is imaged at many positions. Other possibilities could be movement of the detector, beam, or CT stage over the course of the experiment. The gradient was removed empirically by applying a correction to each horizontal slice of the tomogram, dividing each slice by the mean voxel value within a 2D \gls{roi}. The \gls{roi} for the rat specimen was defined as the interior of the solid wall of the semi-annular holder and, for the PMMA phantom, a region of solid material between two of its three cavities.

\begin{figure}
	\includegraphics[width=\columnwidth]{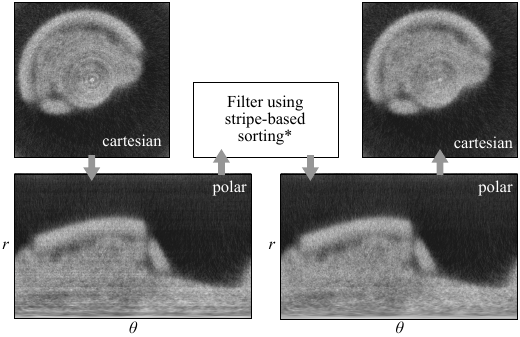}
	\caption{Post-reconstruction ring artifact removal is performed on polar-to-Cartesian mapped versions of each slice of the tomogram. *Algorithm 3 in Vo \textit{et al.} \cite{VoAt18}.}
	\label{fig:ring_artifact}
\end{figure}

Both specimens' resulting tomograms contained visible ring artifacts. We reduced these using post-reconstruction ring filtering, similar to the method introduced by Sijbers and Postnov \cite{SijPo04}. We found this treatment to be more effective than approaches that filter the sinogram prior to reconstruction. Fig. \ref{fig:ring_artifact} shows the method applied to a typical tomogram slice.
The slice is remapped via a Cartesian to polar transform, taking care to ensure that the polar coordinate system origin coincides with the tomogram rotation center. The ring artifacts are thus mapped to stripes, with the unwanted intensity variations now visible along the radial dimension $r$. We applied Vo's stripe-based sorting algorithm (algorithm 3 in \cite{VoAt18}) with a window size of 50, which is a filter originally developed to reduce similar variations in sinograms.
The filtered image was then formed via a final polar-to-Cartesian mapping.

The slices presented in Fig.~\ref{fig:phantom}(c--f) are each the average along the $z$-direction of 61 adjacent tomogram slices, centered around the locations indicated in Fig.~\ref{fig:phantom}(a). This reduced the image noise and uncertainty in image contrast measurements.

\section{Results}
Fig. \ref{fig:phantom}(c--f) shows slices of the PMMA phantom at the positions indicated in \ref{fig:phantom}(a). The same slices in Fig. \ref{fig:phantom}(g--j) reconstructed from transmission data allow us to make comparisons with a common imaging technique that produces high contrast and resolution. All edges of the phantom are clearly visible in the scatter tomogram. The bone-equivalent insert (B) shows up darker than the surrounding PMMA (P) in the scatter tomogram. This is in contrast with the conventional tomogram, which is a map of linear attenuation coefficient, whereas scatter tomography maps electron density. Although self-absorption of photons scattered from (B) makes the values here non quantitative, the reversal of contrast demonstrates that the two imaging techniques display complementary information, and serves as an illustration of the usefulness of scatter imaging. A small difference in value is just visible between the cavities (A) and (B), indicating a greater electron density in (B), as expected.

We quantify image quality using contrast and resolution metrics of the three slices within the phantom in Table~\ref{tab:contrast_resolution}.
\begin{table}
	\caption{Image contrast and resolution measurements}
	\centering	
	\begin{tabular}{rccc}
		& Slice (Fig.~\ref{fig:phantom}) & Contrast & Resolution (\si{\micron})\\
		\hline
		& d & $0.65 \pm 0.06$ & $676 \pm 6$ \\
		Scatter & e & $0.57 \pm 0.06$ & $682 \pm 4$ \\
		& f & $0.49 \pm 0.08$ & $693 \pm 5$ \\
		\\
		& h & $0.85 \pm 0.02$ & $61.6 \pm 0.3$ \\
		Transmission & i & $0.81 \pm 0.03$ & $61.7 \pm 0.3$ \\
		& j & $0.75 \pm 0.04$ & $64.2 \pm 0.4$ \\
	\end{tabular}
	\label{tab:contrast_resolution}
\end{table}
Image contrast $|I_\mathrm{P}-I_\mathrm{A}|/(I_\mathrm{P}+I_\mathrm{A})$ was calculated from the mean value in the PMMA $I_\mathrm{P}$ and in the air-filled cavity $I_\mathrm{A}$, with measurement uncertainty stated in standard deviations. These values were measured from the two circular \gls{roi}s shown in Fig.~\ref{fig:phantom}(b).
The image contrast is greatest in the uppermost reconstructed region for both imaging modalities, due to the background noise raising the mean value of $I_\mathrm{A}$. It achieved a respectable maximum from scatter imaging of $0.65 \pm 0.06$, and a maximum from transmission imaging of $0.85 \pm 0.02$.  

The resolution of the scatter imaging system depends on many factors. The image we see is typically described by convolution of the unblurred tomogram by a \gls{psf}, which is a convolution of the pinhole and \gls{dexel} shape functions, and blurring in the scintillator layer. The \gls{psf} blurs the features in both specimens. In our system, the $0.56 \pm 0.01$\unitspace\si{mm} diameter pinhole is the dominant contributor. A large pinhole was chosen in order to maximize the photon flux reaching the detector, realizing that this would limit the achievable resolution. We measured resolution by azimuthally sampling an annular region of the PMMA phantom straddling the outer circular envelope. A Gaussian test function was then fitted to the gradient of this data, from which the \gls{fwhm} was measured. The highest measured resolution from the scatter imaging was $676 \pm 6$\unitspace\si{\micron}, which is close to the maximum achievable with our pinhole and 50\unitspace\si{\micron} \glspl{dexel}. Our transmission imaging system camera and optics had a pixel size of 16.0\unitspace\si{\micron}, and its measured resolution was $61.6 \pm 0.3$\unitspace\si{\micron}, which is dominated by the scintillator \gls{psf}.

Compton scattered photons have lower energy than the incident X-rays and experience correspondingly greater attenuation in the outgoing scatter path, resulting in self-absorption artifacts when conventional tomographic reconstruction algorithms are used. The artifacts look similar to beam hardening artifacts in polychromatic tomography, and are visible in Fig. \ref{fig:phantom}(c) as an increased value of the substrate insert when compared with the same region in Fig. \ref{fig:phantom}(d). Cupping artifacts are also visible across the extent of the phantom and within the substrate insert. Unlike cupping artifacts in transmission images, which arise due to beam hardening with polychromatic sources, and the addition of unwanted scattered photons \cite{JosSp82}, these artifacts are present in the reconstruction of the scatter signal from the monoenergetic synchrotron source. This is likely due to a combination of the absorption effect of the Compton energy shift described above and the use of a conventional implementation of the inverse Radon transform. That is, the Radon transform is employed validly for a single X-ray source emitting a single ray through the full extent of an object, whereas here the outgoing contribution is the sum of scattering toward the detector from all sites along the path.

Attenuation correction was not attempted and would be required to make quantitative reconstructions from scatter data. It may be possible to substantially correct these artifacts using an approach similar to that of Hogan \textit{et al.} \cite{HogGo91, BruCe02}. Hogan's scheme, which was developed for scatter imaging using pencil beam illumination, is not directly applicable to full-field illumination because the forward scatter imaging model is different. Future extensions of the method described here would aim to develop a suitable correction scheme. In our case, since we expect to capture transmission images along with the scatter data, it would be possible to incorporate them into the correction scheme.

As mentioned earlier, we corrected a residual intensity gradient in the reconstructed tomograms. This boosted the value in the parts of the phantom illuminated by the beam periphery so that it appears equal in all slices (d--f). Because the uncorrected values are closer to the detector noise floor, it also increases the background in those slices that are more strongly boosted. This is increasingly visible in the region outside the cylindrical outer envelope of the phantom in (e) and (f) when compared to (d), and also inside the cavities in (f). The interior of the lower cavity is slightly brighter in (e) and (f), due to the presence of the bone-equivalent insert.

\begin{figure}
	\includegraphics[width=\columnwidth]{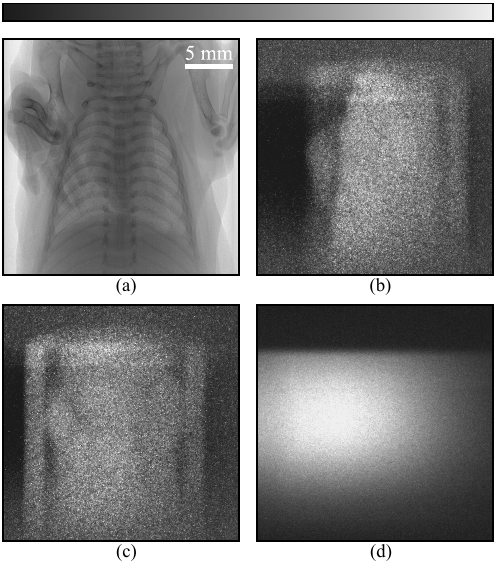}
	\caption{Projection and scatter images of a rat held in a PMMA semi-annular holder, and of a PMMA sheet. (a)~Anteroposterior (AP) transmission image of rat chest at rotation stage angle $\Theta_0$. (b)~Flat-field-corrected scatter image of the rat, taken at the same rotation angle $\Theta_0$. (c)~Flat-field-corrected scatter image of the rat, taken with the stage rotated to $\Theta_0+45^\circ$. (d)~Scatter image of a PMMA sheet used for flat-correction. Image values are in arbitrary units; the colorbars represent values between minimum and maximum limits, with [min, max] ranges for each subfigure as follows: (a) [3, 8], (b) [1, 5], (c) [1, 5], (d) [4, 11].}
	\label{fig:projections}
\end{figure}

Fig.~\ref{fig:projections} shows transmission and scatter images of the rat and PMMA sheet.
Fig.~\ref{fig:projections}(a) is an \gls{ap} transmission image with flat and dark corrections applied, taken at a rotation-stage angle we label $\Theta_0$. 
Image orientation is inverted by the pinhole imaging system, so the scatter images (b--d) have been reinverted here to more easily associate the rat's features with the image in (a), i.e. anatomically oriented with anterior/head at top of image, posterior/tail at bottom.
The scale bar applies only to (a) because the scatter images are significantly distorted by the cone projection and no single length scale applies to all image features.
Fig.~\ref{fig:projections}(b) is the scatter image at $\Theta_0$, after flat and dark correction and despeckling of bright outliers, and (c) is the image after a further rotation to $\Theta_0+45^\circ$.

Fig.~\ref{fig:projections}(d) is the dark-corrected flat-field image at one position of the PMMA sheet. The intensity cuts off at the top due to the beamline's upper horizontal slit blocking the beam.
Because the flat-field image was formed by combining the one shown with a second image taken at a different distance, and hence with different magnification, this resulted in a double image of the slit at that cutoff step. The effect is visible as a narrow horizontal strip of increased intensity in the rat examples (b) and (c) to which flat correction has been applied.
Flat-field correction of these images also introduces noise into those image areas coinciding with low values in the flat-field image; i.e. the top, bottom, and right regions.

The PMMA probe sheet produces a scatter signal from volume regions that are not occupied by the rat or holder. This leads to overcompensation by the flat-fielding scheme, darkening the sides of the image. Attenuation correction of the data will enable the beam profile to be correctly accounted for in the forward model, so that this type of flat-field correction will not be required.

The bright peak of the illuminating beam is visible to the left of the image center in (d). In the second flat-field sheet image (not shown), the PMMA sheet was moved farther from the detector so that the beam's intensity peak shifted to the right, to slightly left of the image center.
As mentioned earlier, the spatial distributions of Compton and Rayleigh scattering differ. Compton scatter dominates at angles beyond about 45${}^\circ$ to the beam axis, with the distribution being energy and element dependent. Because our scatter detector was placed at 45${}^\circ$, we may be observing this transition as a horizontal intensity variation across the flat-field scatter image. This effect may be responsible for the location of the intensity peak lying to the left of center.

\begin{figure}
	\includegraphics[width=\columnwidth]{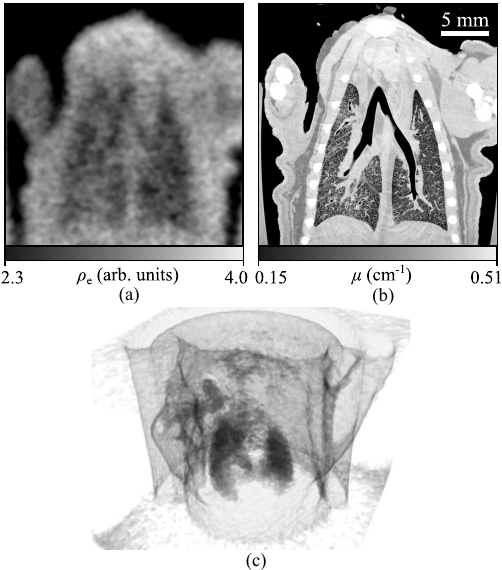}
	\caption{Simultaneously acquired scatter and transmission tomograms of a juvenile rat in a PMMA semi-annular holder. (a)~Coronal image slice from scatter tomogram. Major airways and lung margins, sufficiently clear for segmentation and 3D volume estimation, are visible. (b)~Corresponding image slice from transmission tomogram. (c)~The scatter tomogram shows the lungs and major airways.}
	\label{fig:rat_lungs}
\end{figure}

Fig.~\ref{fig:rat_lungs} shows corresponding coronal slices through the rat tomograms, which were reconstructed from the scatter and transmission images.
The two volumes were aligned and the slice images obtained using 3D Slicer \cite{FedBe12}.
In Fig.~\ref{fig:rat_lungs}(a) the outer silhouette of the rat, some internal tissue-type variation, and the lungs and major bronchi are visible. The contrast at the lung margins should be adequate for segmentation, enabling direct measurement of the lung volume from the tomogram.
The transmission reconstruction in Fig.~\ref{fig:rat_lungs}(b) is provided for comparison. It shows several proximal divisions of bronchi and bronchioles, and lobe boundaries in the lungs, as well as differentiation between tissue types, organs, skeletal features, and fur.
Fig.~\ref{fig:rat_lungs}(c) is a volume rendering of the scatter tomogram with the lungs and major airways clearly visible.

\section{Discussion}
We demonstrated 3D tomographic reconstruction from scattered photons with full-field X-ray illumination of a PMMA phantom and a juvenile rat. In contrast with previous methods, which have used pencil- or fan-beam illumination, this allows the method to be added to existing setups in which full-field transmission imaging is already performed.

There are many avenues for improvement of the imaging system reported here. The resolution is currently limited by the pinhole diameter, which was chosen to match the detector sensitivity, and not aimed at resolving the smallest features in the rat, despite being able to clearly resolve the larger bone-equivalent cylinder in the phantom. This is evident from the inability to resolve the bones, whose diameters are similar to the $\mathcal{O}$(1\unitspace\si{mm}) scatter tomogram resolution achieved.
The resolution was, however, sufficient to resolve anatomical features of the rat airways and lungs, so matching the stated aim of allowing for the possibility of lung volume estimation.

The edges in the phantom appear evenly blurred, indicating an isotropic \gls{psf} and suggesting that deconvolution could be used to enhance detail, albeit at the expense of increased noise. We attempted this but found that the resolution improvement was insufficient to reveal new structure with our specimens.

Any background signal present due to multiple scattering did not prevent us from distinguishing features and different tissues. Indeed, the measured contrast in the scatter slice image was of a similar order to that from transmission imaging. This provides strong evidence for the legitimacy of applying full-field illumination to objects of similar composition and size to ours. Any correction scheme developed in future would further increase the contrast by reducing the cupping artifact visible across the phantom and increasing the measured value in the PMMA substrate.

We now discuss some potential approaches to improving the scatter photon detection efficiency, which apply to both synchrotron radiation and conventional X-ray sources.
Moving to coded apertures or pinhole arrays could increase the resolution and also the photon flux reaching the detector, improving the \gls{snr} and dose efficiency, with further improvement possible by changing the detector technology to direct-detection photon-counting detectors or by optimizing scintillator materials.
The resolution limits of this modality should be the same as for \gls{spect} implemented using a pixelated detector, with the usual tradeoff between spatial resolution and \gls{snr}. The photon emission sources differ between the techniques, but the detection methods are essentially the same. This comparison is useful in helping identify avenues for resolution improvement (see, e.g.\cite{WerAa04}). When pixelated detectors are used, both are limited primarily by the collimation optics and detector pixel size. The X-ray energies used for scatter imaging here permit the use of thinner materials in the construction of collimation optics, potentially providing a resolution advantage when compared with \gls{spect}.

Energy-resolving detectors might provide photon path information via the Compton energy-angle relationship, potentially allowing tomography to be done with less collimation of scattered photons, allowing more to reach the detector.
Because multiply-scattered photons reduce in energy each time they Compton scatter, the impact of background scatter due to multiple-scattering could also be reduced by using energy-resolving detectors to filter out photons with energy below some cutoff value. This would permit higher incident energies necessary for imaging larger or denser objects.
Should this or some other scheme for increasing the energy be developed, application to human medical imaging may come within reach.
Although the increased proportion of multiple-scattering in large objects adds additional complexity to the task of tomographic reconstruction, imaging of thinner extremities of the human body, such as hands, is entirely feasible.

Increasing the number of detectors and combining their measurements would allow a greater proportion of the scatter signal to be captured, thereby improving the sensitivity. This would also provide multiple simultaneous views of the object, opening up the tantalizing possibility of reduced-angle or single-shot tomography. The number of rotation angles required to achieve a given reconstruction fidelity reduces in proportion to the number of simultaneous object views, with a proportionate decrease in overall scan duration and reduction of dose. By single-shot, we mean that enough views of the object are acquired simultaneously so that a \gls{ct} reconstruction is possible without having to reorient the object, giving full volumetric information at a single point in time. It is unlikely that enough angles could be instantaneously acquired to reconstruct a tomogram approaching the same resolution as that achievable using a rotating stage, but resolution sufficient for lung volume estimation might be possible.

Synchrotron radiation is typically highly polarized, with associated minima in the Compton profile for photons scattered horizontally from an object, perpendicular to the beam axis. A scatter detector can be placed at $90^\circ$ but above the object, avoiding the scatter minima. In order to capture the maximally-diverse set of tomographic scatter images for tomogram reconstruction, the object would then be rotated about the horizontal axis. This would introduce difficulties in stabilizing an animal against movement during rotation to acquire a tomographic sequence. The geometry presented in this manuscript is meant to complement an existing geometry in which small animals are imaged with their spines aligned vertically, so as to present the coronal view of the lungs to the transmission detector. Furthermore, placing the detectors forward of the $90^\circ$ horizontal position captures more scattered photons, due to the material scattering-factor profiles for polarized sources. For lab-based sources, the lack of polarization minima offers one advantage of allowing detectors to be placed at $90^\circ$. We used a bright monochromatic synchrotron beam but the method should translate to lab-sources with few changes. In the forward model we presented, a polychromatic cone beam source would replace the monoenergetic parallel beam.

Current commercial clinical \gls{dect} transmission scanners utilize complementary information provided by simultaneous imaging at two energies to produce maps of both attenuation and electron density \cite{AlvMa76}. 
The imaging technique in this paper should also enable the derivation of effective atomic number, as we outline in the Appendix, although this would require registration of the scatter and transmission tomograms, which is avoided with \gls{dect}.
Although current \gls{dect} systems can produce the complementary data without increase in dose compared with single-energy systems \cite{HenFi12, WicHa17}, capture and utilization of the scatter signal provides possible avenues towards further overall dose reduction and reduced-angle or single-shot tomography.

Although attenuation correction is a subject for future work, it will be required to make full-field scatter tomography quantitative. For an overview of this topic, see \cite{BruCe18}. Attenuation correction of the electron density map requires determining and compensating for the attenuation along the photon paths. For single-scattered photons those paths can be determined, then mapped onto the transmission-\gls{ct} linear attenuation map. This requires multimodal image registration of the scatter and transmission tomograms. The relevant attenuation must be recomputed for the scattered photons' (angle dependent) energies. To expose the electron density [see (\ref{eq:scattering_probability})] requires knowledge of the element-dependent scattering cross-section distribution. In the general case, this dictates that a fully-quantitative solution method take an iterative approach. However, for some biological applications such as mapping gold nanoparticle concentrations, or iodine-based contrast agents, where the aim is to map the concentration of a priori known single-elemental content, the scattering cross sections can be more readily determined. When these concentrations are low and the bulk of a sample can be approximated as tissue or water, or when the scatter is through small angles, the correction parameters can be well-approximated, and correction should certainly be possible.

\section*{Acknowledgment}
We thank Martin Donnelley, Brett Williams, Darren Thompson, Katie Lee, Linda Croton, Martin de Jonge, Matthew Dimmock, Michelle Croughan, Andrew Kingston, and Mitzi Klein for helpful discussions related to this work and assistance with experimental preparations. Preliminary experiments relating to this work were performed at the SPring-8 synchrotron BL20B2 beamline with the approval of Japan Synchrotron Radiation Research Institute (JASRI) (Proposal 2017A0132). This research was undertaken on the IMBL beamline at the Australian Synchrotron, part of ANSTO (Proposals M12926, M13447, M14379 and M15223).

\section*{Appendix}
\glsreset{kn}
Here we show that, if the distributions of attenuation and electron density are available, a map of effective atomic number may be derived.
 
The \gls{am} model of X-ray attenuation parameterizes the linear attenuation coefficient's dependence on energy $E$ via \cite{AlvMa76}:
\begin{equation}
	\mu(E) = a_1/E^3 + a_2 f_\mathrm{KN}(E) \equiv a_1/E^3 + \rho_\mathrm{e} \sigma_\mathrm{KN}(E),
	\label{eq:Alv-Mack}
\end{equation}
where $f_\mathrm{KN}(E)$ is the \gls{kn} function, $\sigma_\mathrm{KN}(E)$ is the total \gls{kn} scattering cross section, and $\rho_\mathrm{e}$ is electron density. Here $a_1 \approx K_1\rho Z^4/A$ and $a_2 \approx K_2 \rho Z/A$, where $K_1$ and $K_2$ are tabulated constants, $\rho$ is mass density, $Z$ is the atomic number, and $A$ is the atomic mass number.

The term $\rho_\mathrm{e} \sigma_\mathrm{KN}(E)$ in the \gls{am} model does not explicitly include Rayleigh scattering, as we do in our model [see (\ref{eq:scattering_probability})]. We continue the description here according to (\ref{eq:Alv-Mack}), noting that implementation should consider this difference.

For low-Z elements up to around Z = 27 (calcium), the atomic mass number $A \approx 2 Z$.
From this, the mass density may be related to electron density via the atomic mass constant $m_\mathrm{u}$ according to
$\rho = 2 m_\mathrm{u} \rho_\mathrm{e}$. For a compound, $Z$ can be approximated as an effective atomic number $Z_\mathrm{eff}$.

A conventional \gls{ct} reconstruction provides a 3D map of $\mu(E)$, whereas an attenuation-corrected scatter map provides a 3D distribution of $\rho_\mathrm{e}$, or alternatively $a_2 f_\mathrm{KN}(E) \equiv \rho_\mathrm{e} \sigma_\mathrm{KN}(E)$. Regardless of which scatter distribution is available, $f_\mathrm{KN}(E)$ and $\sigma_\mathrm{KN}(E)$ are both well-known \cite{AlvMa76}, enabling recovery of $\rho_\mathrm{e}$ and $a_2$.
Having both $\mu(E)$ and $a_2$ (or $\rho_\mathrm{e}$) allows us to recover $a_1$ via (\ref{eq:Alv-Mack}):
\begin{equation}
    a_1 = E^3 [\mu(E) - a_2 f_\mathrm{KN}(E)] = K_1 \rho_\mathrm{e} m_\mathrm{u} Z_\mathrm{eff}^3.
\end{equation}
Finally, this enables us to recover the map of the object's effective atomic number
\begin{equation}
    Z_\mathrm{eff} = \left( \dfrac{a_1}{K_1 \rho_\mathrm{e} m_\mathrm{u}} \right)^{1/3}.
\end{equation}



\bibliographystyle{IEEEtran}

\bibliography{tmi}



\end{document}